\documentclass[12pt,preprint]{aastex}

\newcommand\etal{{\it et al.~}}

\newcommand\eg{{\it e.g.,~}}

\shorttitle{MHD Simulations of Relic Radio Bubbles}
\shortauthors{De Young & Jones}

\begin{document}

\title{MHD Simulations of Relic Radio Bubbles in Clusters}


\author{T. W. Jones}
\affil{Department of Astronomy, University of Minnesota,
 Minneapolis, MN 55455}
\email{twj@astro.umn.edu}

\author{D. S. De Young\altaffilmark{1}}
\affil{National Optical Astronomy Observatories, Tucson, AZ 85719}
\email{deyoung@noao.edu}


\altaffiltext{1}{ NOAO is operated by AURA, Inc.\ under contract to the
National Science Foundation.}


\begin{abstract}

In order to better understand the origin and evolution of relic
radio bubbles in clusters of galaxies, 
we report on an extensive set of 2D MHD simulations of hot buoyant
bubbles evolving in a realistic intracluster medium. Our
bubbles are inflated near the base of the ICM over a finite time 
interval from
a region whose magnetic field is isolated from the ICM. We confirm 
both the early conjecture from linear analysis and the later 
results based on preformed MHD bubbles; namely, that very modest 
ICM magnetic fields can
stabilize the rising bubbles against disruption by 
Rayleigh-Taylor and Kelvin-Helmholtz instabilities. We find 
in addition that amplification of the ambient fields as they
stretch around the bubbles can be sufficient to protect
the bubbles or their initial fragments even if the fields are
initially much too weak to play a significant role early
in the evolution of the bubbles. Indeed, even with initial
fields less than a micro-Gauss and values of $\beta = P_g/P_b$ 
approaching $10^5$, magnetic stresses in our simulations eventually 
became large enough 
to influence the bubble evolution. Magnetic field influence
also depends significantly on the geometry of the ICM field and on
the topology of the field at the bubble/ICM interface. For example,
reconnection of anti-parallel fields across the bubble top greatly
reduced the ability  of the magnetic field to inhibit disruptive
instabilities. Our results confirm earlier estimates of $10^8$
yr for relic radio bubble lifetimes and show that magnetic fields can
account for the long term stability of these objects against disruption
by surface instabilities.  In addition these calculations show that
lifting and mixing of the ambient ICM may be a critical function of
field geometries in both the ICM and in the bubble interior.

\end{abstract}



\keywords{galaxy clusters -- radio sources -- MHD}


\section{Introduction and Motivation}

Recent combined x-ray and radio observations of 
galaxy clusters 
have revealed a growing number of instances in which there is evidence
on scales $\gtrsim 10$ kpc that extended radio sources produced by active galaxies in the cluster clearly
displace the hot ambient intracluster medium (ICM), producing radio-filled,
X-ray holes or ``radio bubbles'' (\eg B\"ohringer \etal 1993,
Fabian \etal 2000, McNamara \etal 2001, Nulsen \etal 2002, Taylor \etal 2002, Blanton \etal 2004). 
Occasionally, as in the Perseus cluster, larger, apparently free
floating radio quiet bubbles or ``ghosts'' are seen outside the inner radio 
structures (\eg Fabian \etal 2003).
In several clusters an additional, remarkable phenomenon is
observed; namely, apparent ``relic'' extended radio structures some tens of kpc
from the central AGN with no apparent or at most faint radio connections between 
them (\eg Fujita \etal 2002, Mazzotta \etal 2002, Young \etal 2004).
Together these various phenomena seem to infer that outflows from AGNs 
can inflate bubbles in the ICM, which then may dynamically evolve
on their own during subsequent periods of reduced or absent energy
outflow from the AGN. Recent observational summaries of these objects
are given in Kempner \etal (2003) and Birzan \etal (2004). For simplicity we refer below to
these isolated objects simply as relics
\footnote{We do not
address here another class of diffuse, ``relic'' radio sources
in clusters, such as A3667, that
may depend in part on large-scale shocks in current merger 
activity (\eg Roettiger \etal 1999) and that have recently been termed
``radio gischt'', or ``radio froth'' (Kempner \etal 2003).} or bubbles.

In general the relic radio sources, and their more active inner companions,
are smaller than classical FR-II objects, yet they often do not have the morphology
that is characteristic of the lower powered FR-I radio sources (McNamara 2002).
Typical radii of the relic radio bubbles are about 10 kpc, and their distances from
the center of the parent galaxy range from about 20 kpc to 50 kpc. Thus these objects
often lie in a transition region between the circum-galactic environment and the actual
intracluster medium (ICM), a condition that will be addressed in more detail in \S 2.
The radio luminosities of the relic bubbles lie in the lower end of that seen in 
the FR-I radio sources, while the luminosities of the inner components,
when present, are comparable to those of FR-I objects. 

A critical derived parameter
for these objects is their estimated age. Age estimates for extended radio sources are
notoriously uncertain (e.g., De Young 2002), but in the case of the radio bubbles an age
estimate can be obtained independently of the usual synchrotron lifetime or Doppler boosting arguments. 
Because the radio bubbles displace the surrounding hot x-ray emitting gas, it can be 
argued that the density inside the bubbles is less than that of the surrounding medium,
but that the interior gas is very energetic and in pressure equilibrium with the 
ambient gas. Under these conditions the bubbles will rise in the cluster gas as
buoyant objects, and one can calculate the buoyant rise times of these objects from a
point nearer the galaxy where the energetic material may have been injected. Such
calculations have been done by e.g., Churazov et al 2001, 
and the resulting rise times are of order $10^{8}$ yr.
Similar results have been obtained by Birzan et al. (2004) from an examination
of the observational data. 
While these calculations are also based upon several assumptions, it is interesting
that the resulting ages for the relic bubbles are comparable to age estimates for radio
sources obtained through the usual synchrotron aging arguments. However, if the radio
bubbles are in pressure equilibrium with their surroundings, their internal energy
densities exceed the synchrotron equipartition energy densities by about a factor of 
ten or more.  
This may have important consequences for the nature of the particle content of
the radio bubbles and possible re-acceleration mechanisms (De Young 2003).

Given the above estimates for the parameters of the radio bubbles, a key issue immediately
emerges.  A clear observational result is that these objects are intact entities.  Yet
buoyant bubbles rising through a surrounding medium are unstable to fragmentation if the
Reynolds numbers are at all large.  This clearly should be the case here if hydrodynamic
behavior is followed, since the hydrodynamic Reynolds numbers in this case are extremely large.
Hence, the rising bubbles should quickly fragment due to both the Rayleigh-Taylor (R-T) and
Kelvin-Helmholtz (K-H) instabilities along the top and sides of the bubble. Detailed
numerical calculations have shown that this is in fact the case (e.g., Churazov et al. 2001,
Br\"uggen et al. 2002). In particular,the detailed high-resolution simulations of Br\"uggen \& Kaiser (2002)
have shown that buoyant bubbles under conditions appropriate to the relic radio sources
will become unstable and fragment into complex substructures in times of $\sim 10^{7}$ yr,
which is much less than the estimated age of the observed relic radio bubbles.  Hence in
order to account for the observed data, some stabilization mechanism must be at work, and
an obvious candidate is the intracluster magnetic field, perhaps in concert
with the internal bubble magnetic field.  

It has been known for some time
that the hot gas in clusters of galaxies often has within it a significant magnetic field
(e.g., Carilli \& Taylor 2002, Taylor et al. 2002), with estimated typical magnetic field strengths
of $\sim 5 \times 10^{-6}$G ($5~\mu$G). 
It is also well known that magnetic fields can stabilize both
the K-H and R-T instabilities (e.g, Chandrasekhar 1961; Shore 1992) if there is a significant
field component that is parallel to the interface between the two fluids.  The expansion of
a radio source into an ambient medium that contains a tangled magnetic field will ensure that
such tangential field components exist at the bubble-ICM interface, independent of any magnetic
field that is contained within the radio source itself.  This stabilizing effect of intracluster
magnetic fields was pointed out by De Young (2003) who derived analytic conditions for the
stabilization of the relic radio bubbles in the hot ICM. Those calculations showed that the
field strengths observed in many clusters will in fact stabilize the bubble interface in the
linear regime, and hence the cluster magnetic fields could account for why the relic radio bubbles
are seen as intact objects at such late times. Additional support for this idea is found in the
two-dimensional MHD calculations of Br\"uggen and Kaiser (2001), who 
applied a $\beta \sim 10$ field inside the bubbles that was aligned with
the bubble surface.
Those calculations treated a scale
much larger than that of the relic radio bubbles, covering regions 200 to 400 kpc in extent. 
Despite the large scales and the rather strong magnetic fields
employed, the results are suggestive in that the radio
source interface may be stabilized, and their resulting geometry is similar to that seen in the relic
bubbles in clusters.

A more recent two-dimensional MHD study has been performed by Robinson \etal (2004) (hereafter R04). That
work also addressed the evolution of relic radio sources in clusters, and it reported
the results of both hydrodynamic and MHD calculations. The hydrodynamic results are very similar
to those reported earlier, with the early onset of instability and subsequent fragmentation of the
relic radio source bubbles. A basic result from R04 in the MHD case is that again the
presence of even relatively weak magnetic fields can serve to stabilize the relic radio lobes as they rise in the ICM,
which serves to verify the earlier considerations of magnetic effects mentioned above.
The calculations of R04 were limited to two initial magnetic field
configurations. One was a uniform 
magnetic field threading both the bubble and the ambient
medium, the orientation of which was placed along each of the three 
principal axes in successive simulations. The other configuration
involved a bubble supported against external gas pressure
by an internal by force-free field. The ambient magnetic field
in the computational plane was set to zero in that model, although it included
a uniform field transverse to the computational plane that
contributed (only) to the total pressure.
Finally, R04 initialize their calculations with the bubble already in place, and assume a reflection symmetry across the
vertical mid-line of the plane. 

The new results described below are extensions of the work by R04 that are physically
significant and immediately relevant to the evolution of relic radio bubbles in clusters of galaxies.
They are still two dimensional. Like R04 we defer three 
dimensional simulations to follow-up work, since many 2D dynamical
features remain qualitatively valid in
3D flows but are much easier to isolate and analyze in the 2D case .
In contrast to the R04 simulations, the present calculations 
consider the effects of varying not only the ambient magnetic field
orientation, but also its strength as a function of position. This provides
a broader understanding of the roles played by the external
magnetic field, permitting, for example, the simulation of cluster
atmospheres with constant $\beta = P_g/(8\pi B^{2})=nkT/(8\pi B^{2})$, which may be dynamically more likely in the
intracluster environment. We also created our more general initial
field configurations so that the bubble and external fields
were isolated from one another, reflecting the distinct origins of the
two fields. We also varied the topology of the field
along the bubble/ICM interface, since that influences the
magnetic reconnection expected there. In addition, while some of the 
R04 simulations were full MHD, those authors chose not to describe
the behaviors of the magnetic fields in and around their bubbles,
with the special exception of the one magnetically supported bubble. 
In addition, because the distance from the center of the parent galaxy to the
relic radio bubbles is only a few tens of kiloparsecs, a realistic equilibrium ICM/ISM configuration
is necessary that incorporates both the effect of the galaxy gravitational potential and that due to
the inner regions of the cluster itself. Third, and perhaps most 
significantly, the present calculations differ from those of
R04 in that we 
followed the dynamical evolution of radio bubbles as they were inflated and then evolve away from the source that
supplies their internal energy. This is an essential element in determining realistic evolutionary
tracks for the relic radio bubbles. 
The internal magnetic fields of the radio bubbles
are evolved in a self consistent manner as the bubbles are inflated 
and then ``cast off'' into the ICM.
Parameters for both the ambient ICM and the inflating bubbles are varied over a wide
range in order to best determine which conditions may be most applicable to the cluster relic radio
sources.

The plan of the paper is as follows. Section 2 describes the method of calculation used for these
MHD simulations and a description of the initial and boundary conditions for
the various cases considered. The results are presented in Section 3, together with an analysis of
the physical processes at work.  A summary and conclusions are presented in Section 4.

\section{Calculation Details}

Our simulations assumed 2D Cartesian geometry on the
$y-z$ plane, with nonuniform gravity in the $-z$ direction, as outlined below. 
The computational domain was a square, 55 kpc on a side, 
spanning [5,60]kpc in $z$ and centered on $y = 0$.
Since radiative cooling is not important
over the times simulated (see R04), we ignore that process and assign
an adiabatic index, $\gamma = \slantfrac{5}{3}$, to
the compressible, conducting fluid.
Further details of the ambient ICM are given below.
The bubbles were inflated from a region of circular cross-section with
radius $r_b = 2$kpc, centered at $y \approx 0$, $z = 15$kpc. 
Much of our study focuses on the evolution of instabilities
that developed from irregularities on the surfaces of the expanding 
bubbles. Discrete mapping of the circular inflation region onto
the Cartesian grid automatically creates seed perturbations for
the instabilities. However, the simplest mapping, with the inflation
region centered at the intersection of grid-zone boundaries, introduces
an artificial mirror symmetry into the simulation. To break that symmetry,
and to ``naturalize'' the seed perturbations as much as possible,
we applied two small modifications to the initial conditions.
First, we offset the center of the inflation region in the $y$ coordinate
by $0.25 \Delta y$, where $\Delta y = \Delta z$ is the size of a numerical zone.
In addition, we added a 1\% random noise to the ambient density
distribution, providing a second, independent seed contribution of
comparable amplitude to the mapping discretization. This noise was
too small to have any other significant influence on the simulations.
Other bubble creation details are presented below. 

We used our well-tested 
ideal, compressible TVD ideal MHD code described in Ryu \etal 1998. The
scheme is second order accurate in both space and time and is designed to
capture cleanly all families of MHD discontinuities. It
employs a constrained transport technique for updating the
magnetic field, so that the initial divergence-free field condition is
maintained to machine accuracy. This version of the code is 2\slantfrac{1}{2}
D, so that all three components of vector fields
are included. However, the computations reported here all have
$B_x = u_x = 0$. Both open $z$ boundaries were designed to maintain
hydrostatic equilibrium; the two $y$ boundaries were open, as well. To
track the bubbles we employed a passively advected mass-fraction tracer assigned
value $C_f = 1$ for material originating inside the bubble inflation
region, with $C_f = 0$ elsewhere. Simulations were carried out
on grids ranging from $256^2$ to $1024^2$. Only the two higher
resolution cases are discussed here. Those resolutions are
comparable to or exceed the stated effective AMR MHD resolutions of
R04.

We add some brief general comments about considerations
related to the finite numerical resolution involved in these simulations.
Dissipation in our ideal MHD code results, of course, from approximation
and roundoff errors rather than numerical models for
those effects. Viscous and resistive dissipation are, thus,
dependent on numerical resolution. Tests show that the
effective kinetic and magnetic Reynolds numbers scale
as $R \propto l^k$, where $l$ is the scale of interest and
$k$ lies between 1.5 and 2, thus mimicking so-called ``hyperviscosity''
and ``hyperresistivity'' rather than normal 
viscosity and resistivity (\eg Ryu etal 1995, Lee \etal 2003).
On the other hand, while the physical viscosity and resistivity in
collisionless astrophysical plasmas are not well-determined, they
are unlikely to be expressible as simple, ``normal-type'' dissipation
constants.
In all the models run we found the results to be
qualitatively independent of the three resolution used, since
in all cases the effective kinetic and magnetic Reynolds numbers
on scales that dominate the dynamics (more than a few zones)
exceed several hundred.
That consistency agrees with our earlier resolution studies with this
code (\eg Jones \etal 1996, Miniati \etal 1999). While
the fine details of the observed behaviors do depend on numerical
resolution for several reasons, our aims here are more general and
not significantly impacted by resolution effects. A specific note
about magnetic reconnection is also warranted in this regard. Reconnection
is a topological transformation of the magnetic field that requires
dissipation in order to take place. While details of that transformation
certainly depend on the details of the dissipation, previous
2D and 3D numerical studies with this code have found that global 
behaviors are once
again consistently obtained once the magnetic Reynolds numbers
become large (\eg Jones \etal 1997, Ryu \etal 2000).

\subsection{The Gravitational Potential and Ambient Medium}

Because the observed relic radio bubbles lie $\sim 20 - 50$ kpc from the
nuclei of the parent galaxies, the buoyant bubbles of interest rise in an
environment that is influenced by both the mass contained in the
cluster core and that of the parent galaxy. Hence, the gravitational
potential used in our simulations is a superposition of those due to 
both the cluster and the active galaxy. Observational data that supply
kinematic verification of such models are rather scarce, but this problem 
has been treated in a different context by Kelson et al. (2002). The
present calculation follows their approach. The mass distribution
in the cluster core, including dark matter, follows an NFW model
(Navarro, Frenk \& White 1997) with a total density distribution given
by $\rho_{c}(\zeta_c) \propto 1/[\zeta_c^{\alpha} (1 + \zeta_c)^{3 - \alpha}]$, where
$\zeta_c$ is a normalized cluster radius; $\zeta_c = r/r_{s}$. Several choices are
available for modeling the mass distribution of the parent galaxy,
such as an $r^{1/4}$ law, a ``King'' model (King 1966, 1978), or another
power law distribution. Kelson et al. found that the observational
data in the inner regions were relatively insensitive to the model
chosen for the galaxy mass distribution, and hence we chose an
easily integrated King model
of the form $\rho_{g}(\zeta_g) = \rho_{o}/[1 + \zeta_g^{2}]^{3/2}$, 
where $\zeta_g = r/a$. This gives
$M(\zeta_g) = 3 M_{c} [\ln(\zeta_g + \sqrt{1 + \zeta_g^{2}}) - \zeta_g/
\sqrt{1 + \zeta_g^{2}}]$, where, $a$ is the core radius of the 
galaxy and $M_{c}$ is the core mass. 

\begin{figure}
\epsscale{.90}
\plotone{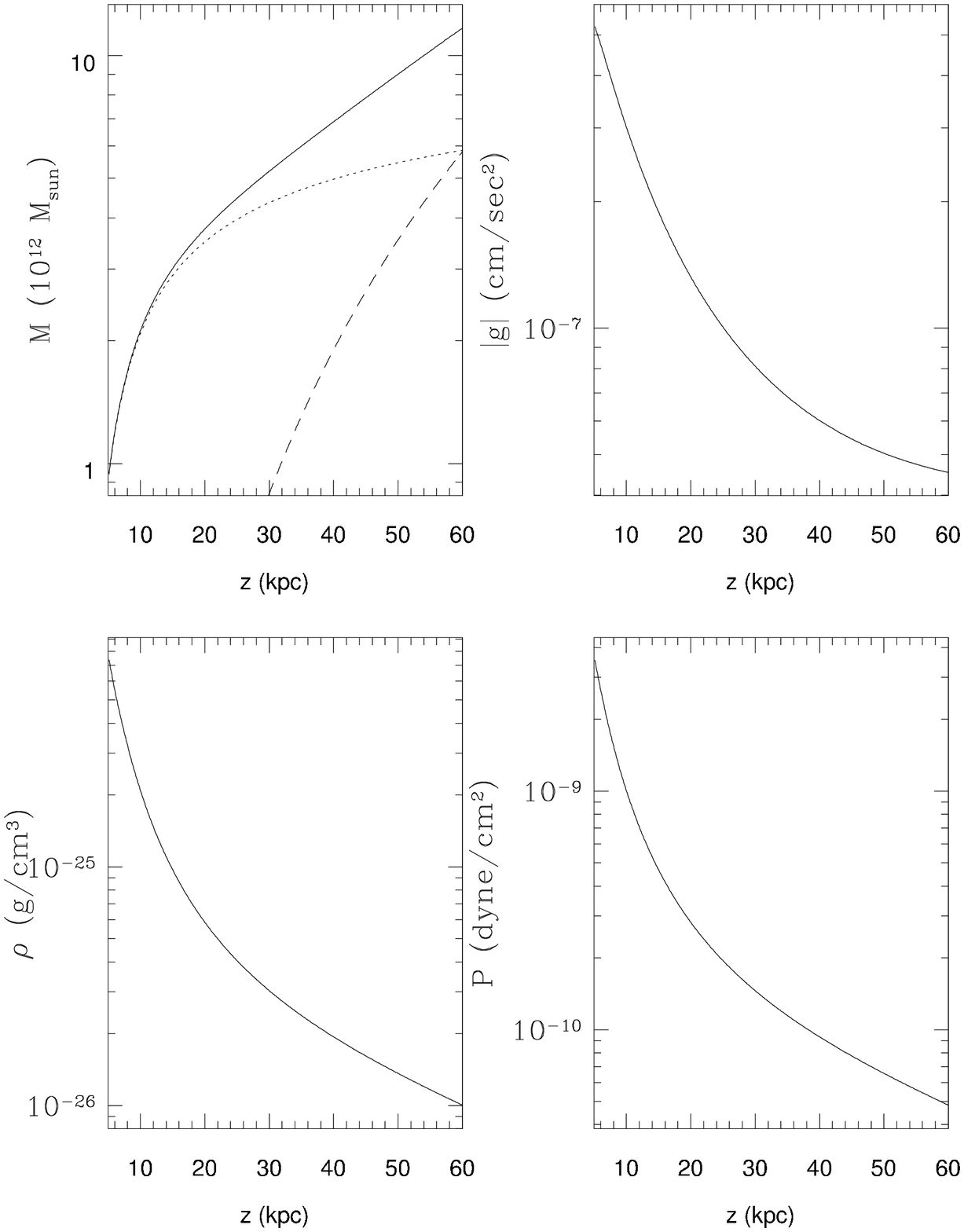}
\caption{Initial environment of the bubble simulations. 
Top left: Gravitational mass as
a function of radius from the core -- total (solid), galaxy (dotted),
cluster (dashed). Top right: gravitational acceleration, Lower left:
gas density, Lower right: gas pressure. Distance from the mass center
is represented by $z$ to reflect the symmetry of the MHD simulations.}
\label{setup}
\end{figure}

In our simulations we assumed
an NFW cluster model with $\alpha = 0$, $r_{s} = 400$kpc,
and a normalization of $M_{cluster} = 3.5\times 10^{10}{\rm M}_{\sun}$ at
10 kpc. This yields a cluster mass of $3.5\times 10^{12}{\rm M}_{\sun}$
at 50 kpc. For the galaxy, a core radius of $a = 3$kpc was chosen,
normalized to a galaxy mass of $3.5\times 10^{12}{\rm M}_{\sun}$ at a radius
of 20 kpc. With this model, the equal mass contribution crossover point
occurs at a radius of about 60 kpc. The total and constituent mass
distribution, along with the radial gravitational acceleration are shown in Figure \ref{setup}.
Since our simulations were carried out on a Cartesian grid,
we substituted the $z$ coordinate for $r$, and will henceforth use
$z$ to indicate vertical distance. 

Again applying a Cartesian geometry, we took the ICM
to be an isothermal plane stratified 
hydrostatic atmosphere with
$kT = 3 {\rm keV}$ ($T \approx 3.5\times 10^7 {\rm K}$) and 
a hydrogen mass fraction $X = 0.75$ 
(mean molecular weight, $\mu \approx 0.6$), yielding an ICM adiabatic
sound speed $c_{sI} = \sqrt(\gamma P/\rho) = 0.914~{\rm kpc/Myr} = 894 {\rm km/sec}$. 
(Henceforth, we shall use the subscript ``I'' to indicate ICM
initial conditions.)
An electron density in the
atmosphere set to $n_e = 0.1 {\rm cm}^{-3}$ at $z = 5$ kpc
leads to the density and pressure distributions shown in Figure \ref{setup}.
The ICM scale height, which we can define to be $h = c^2_{sI}/|g|$,
increases roughly linearly from $h \approx 5~{\rm kpc}$ at the bottom 
of our computational domain to $h = 57~{\rm kpc}$ at the top. 
That is, $h \approx z$.
This behavior allows us to derive a simple and useful form for
the variation of the pressure and density with height. In fact, it is
easy to show that $P \approx  P_0 (z/z_0)^{-\gamma}$, with a similar
form for density. We have verified this to be a good fit to the actual
ICM profiles.

The initial magnetic fields of the ICM and the bubble
were isolated from one another. The ambient field was
generated from a magnetic stream function, $A_x$, given in polar
coordinates with respect to the bubble injection center as
$A_x = B_0 \sin{(\phi-\phi_0})(1 - (r_b/r)^2)$ for $r > r_b$, 
where $\phi$ and $\phi_0$ are measured from the $y$-direction, 
and $r_b$ is the radius of the bubble inflation region.  $\phi_0$
rotates the field structure around the bubble injection center, to
allow an arbitrary orientation.
The magnetic field components are $B_r = B_0 \cos{(\phi-\phi_0})(1 - (r_b/r)^2)$
and $B_{\phi} = -B_0 \sin{(\phi-\phi_0})(1 + (r_b/r)^2)$. 
For $\phi_0 = 0$ this becomes a uniform horizontal 
field ($B_z = 0,~B_y = B_0$) as $r \rightarrow \infty$. As $r \rightarrow  r_b+$,
$B_r  \rightarrow 0$, so that the field becomes tangential to
the bubble inflation region, with a peak magnitude equal to $2B_0$. 
Initial ambient fields defined by $A_x$
alone are identified below as ``uniform'' or ``U'', since away from the
bubble they are nearly so.
For the simulations discussed below $B_0 > 0$. We carried out ``uniform''
field simulations with $\phi_0 = 0\degr, -45\degr, -90\degr$, although
we omit discussion of uniform fields with $\phi_0 = 0\degr$ and
$\phi_0 = -90\degr$, since
they are very close in behavior to the analogous models described
in R04.

In addition we computed $\phi_0 = 0$ models, designated below as
``constant $\beta_{I}$'' or ``B''. There we modified the magnetic field
found from $A_x$ and
the gas pressure illustrated in Figure \ref{setup} so 
that $P(z) \rightarrow P(z)\beta_I /(1+\beta_I)$
with $2P/B^2_y(y \rightarrow \infty) = \beta_I$, where $P$ is the gas 
pressure. This maintains hydrostatic equilibrium in the $z$-direction
with an almost constant $\beta_{I}$. 
By rescaling the ``U'' model $B_y$ in proportion to $\sqrt{P(z)}$ and
applying a small correction to $B_z$ it is possible in this model
to maintain the divergence-free condition in the initial ICM field. 
The smallest $\beta_I$ modeled
was 120, so equilibrium in the $y$-direction was not significantly
disturbed by this procedure. 

\subsection{Bubble Inflation and Magnetic Field}

At the start of each simulation the ``bubble inflation'' region was established 
and maintained for a time $t_i$ that ranged over $0 \le t_i \le t_e$,
where $t_e$ was the end time of the 
simulation. The bubble inflation region had a
radius $r_b = 2$kpc, centered at $z$ = 15 kpc, and it was given 
a density equal to 1\% of the ambient density 
of the ICM at $z$ = 15 kpc. We gave the inflation region 
constant pressure, matching the ambient pressure at $z = 15 {\rm kpc}$.
The associated sound speed internal to the bubble, $c_{sb}$, 
consequently exceeded that of the ICM by a factor of ten.
The uniform pressure inside the inflation region resulted in 
a slight, 10\% gas overpressure compared to that of the ambient medium on
the upper boundary of the inflation region, which led to a brief,
dynamically insignificant spike in the initial energy flux into the bubbles.
The inflating bubbles expanded subsonically with a density
comparable to the inflation region. They were
lifted upward by buoyancy with a velocity
$\sim 0.2-0.6 c_{sI}$ as discussed in \S 3.1. The bubble inflation power,
defined as the total energy flux across the boundary of the inflation
region, is given by 
$\dot E_b =  \ointop ds~u_i(\slantfrac{5}{2}P_i + P_{bi}  + \slantfrac{1}{2}\rho_i u^2_i)$,
where $P_i$, $P_{bi}$, $\rho_i$ and $u_i$ represent the gas pressure,
magnetic pressure, gas density and radial outflow velocity at the boundary of 
the inflation region, respectively.
Since the pressure and density inside the inflation region
remained fixed during the inflation
interval, $\dot E_b$ was controlled by $u_i$, which was always small
compared to the internal bubble sound speed, $c_{sb}$. Therefore, since
the Poynting flux is also generally negligible,
$\dot E_b \approx 2\pi r_b u_i \slantfrac{5}{2}P_I$.
Outflow speeds into the inflating
bubbles mirrored, in fact, the upward buoyant bubble expansion.
Thus, since this leads to $u_i \lesssim 0.5 c_{sI} \lesssim 0.05 c_{sb}$,
$\dot E_b$ typically increased at the start of the simulations towards 
$\sim 10$\% of the characteristic value,
$\dot E_{0} = 2\pi r_b P_I c_{sb}$. If the inflation region
was maintained longer than the time required for the inflating bubble to
rise through one ICM scale height ($\propto h/c_{sI}$), typically 
about 20 Myr in these simulations, a crude de Laval nozzle formed
about one scale height above the inflation region. 
In such cases a collimated, transonic to mildly supersonic flow continued higher
into the ICM so long as the inflation was maintained. That plume was capped by a
bubble, analogous to the flows simulated by Br\"uggen \& Kaiser (2002).

A circumferential magnetic field was imposed inside the
active inflation region, given by $B_r = 0$,
$B_{\phi} = B_1(r/r^{'}_b)$, for $r \le r^{'}_b=r_b -4\Delta z$, 
then decreased quadratically to zero at $r = r_b$ in order to isolate
the bubble magnetic field, initially.
In the models discussed $|B_1| = 2 |B_0|$, in order to match the peak
perimeter field in the ``U'' models. 
The same ratio was used in the ``B'' models. In that case, the
ambient field strength varies asymmetrically across the bubble,
so there is no simple match in the internal and external field
strengths. On the other hand, in both types of models, the
magnetic field inside the bubble is strongly modified by flow,
especially shear, so these details are not very significant.
During the inflation period, magnetic fields advected from the inflation
region were lifted into the rising bubble and stretched
around the interior perimeter, enhancing the magnetic field just inside
portions of the side and bottom boundaries by as much as a factor of two,
which is qualitatively an expected result. 
In most cases $B_1$ was positive, giving a
counterclockwise sense to the initial bubble field. Then the internal
and ambient fields were parallel at the bottom of the bubble and
anti-parallel at the top when $\phi_0 = 0\degr$. 
For the two cases identified below by the
letter ``R'', the internal field was reversed with respect to this
convention. 

We note that the initial magnetic field configurations are 
divergence free everywhere. This is clearly true inside the bubble inflation
region and outside, as well. A little thought shows that it is also true
across the boundary between them. The key point is that no magnetic
flux crosses that boundary. Thus, an accounting of the flux through
a cylindrical test shell including that boundary can be separated into
two independent parts with the boundary dividing them.  Each piece
separately has zero net flux, so the full test box does, as well.

\section{Discussion}

We carried out a total of 28 model simulations, varying the magnetic
field geometry and strength, its topology along the bubble boundary,
the time for which bubble inflation was
maintained, as well as the numerical resolutions used. We included
models with ``preformed'' bubbles ($t_i = 0$) in uniform
ambient magnetic fields that were both horizontal ($\phi_0 = 0\degr$)
and vertical ($\phi_0 = -90\degr$) in order to mimic the
kinds of models studied in R04.
Those simulation behaviors resemble qualitatively the results reported in 
R04 for similar bubbles,
confirming their results. We will not discuss them here.
Rather, in order to extend our understanding, we focus instead on ten
representative simulations that involved nonuniform horizontal or 
uniform oblique ambient fields ($\phi_0 = -45\degr$), with
finite periods for inflation of the buoyant bubble plasma. 
Table \ref{tbl-1} summarizes their properties.

\subsection{Bubble Energetics and Buoyant Dynamics}

\begin{figure}
\plotone{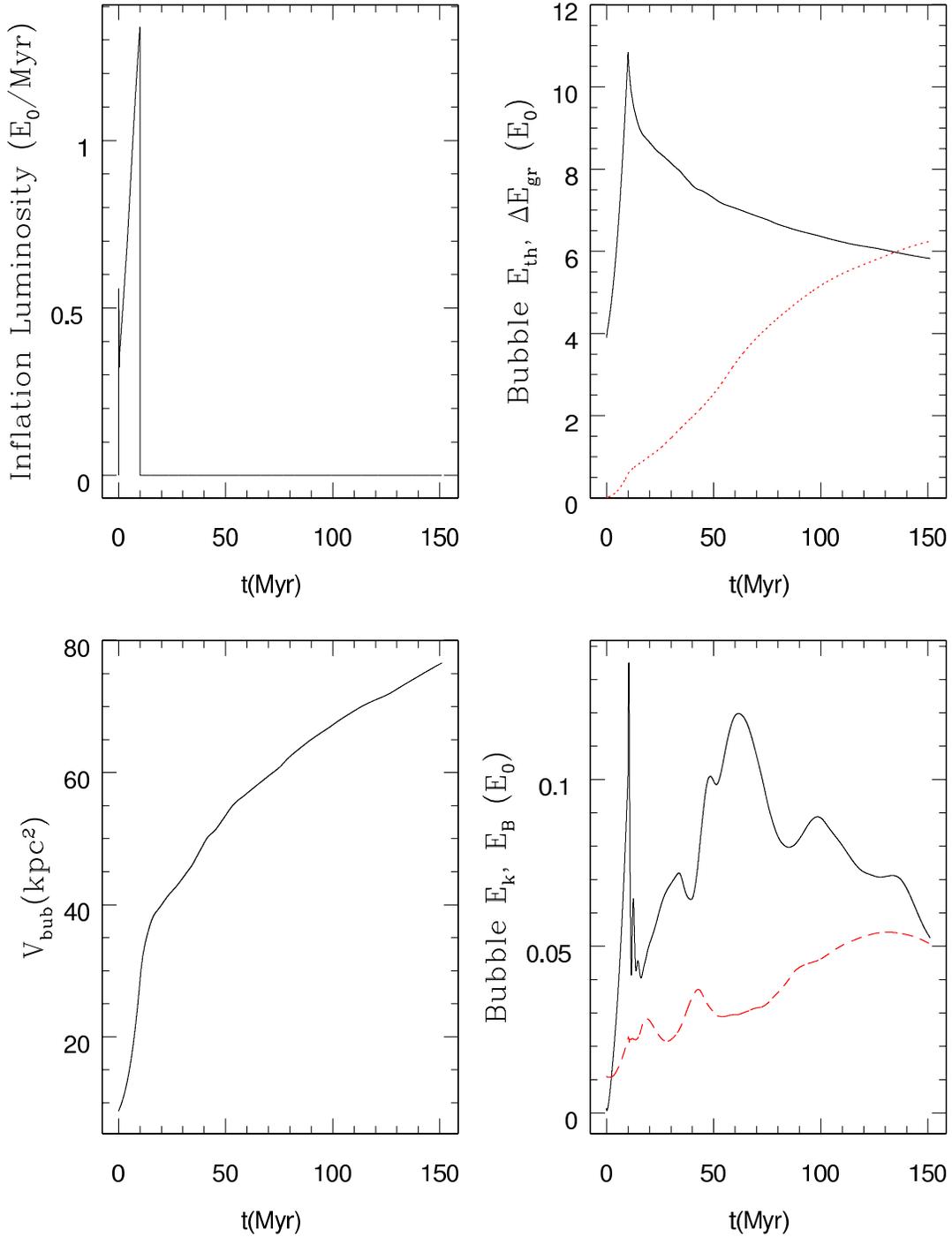}
\caption{Energetics of the bubble in model
BM$_h$. Top: left- Total 2D energy flux through the boundary of the
bubble inflation region; right- Bubble thermal energy (solid line) and
net change in gravitational energy (dotted line). Bottom: left-Bubble 2D ``volume'';
right- Bubble kinetic energy (solid line) and magnetic energy (dashed line). 
Energy units are $E_0 = 5.5\times 10^{55}{\rm erg}$.}
\label{bubstat1}
\end{figure}

\begin{figure}
\epsscale{.85}
\plotone{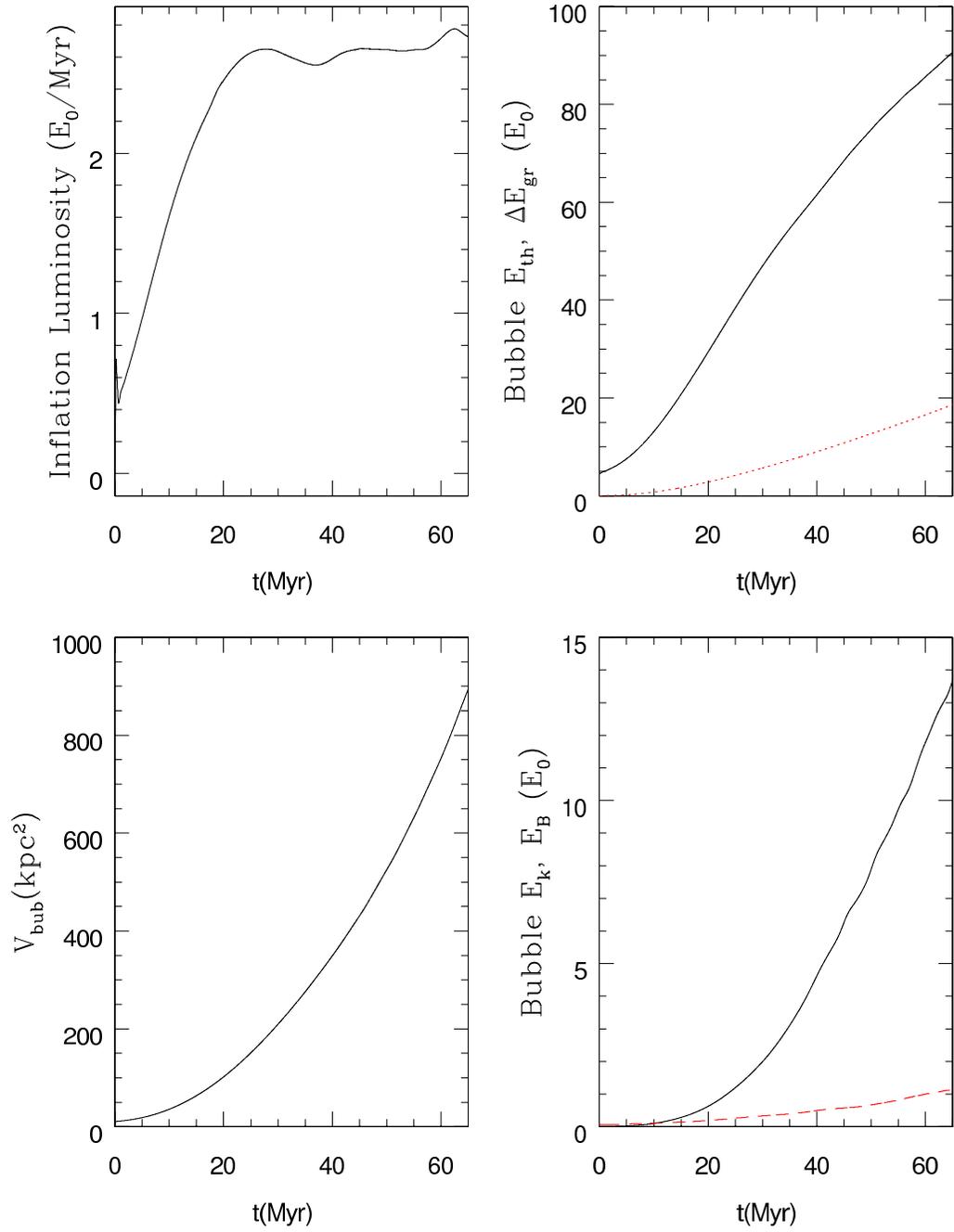}
\caption{Same as Figure \ref{bubstat1}, except for model BS-C.}
\label{bubstat5c}
\end{figure}

Our main objective is to understand better the roles of magnetic fields
in the bubble dynamics. However, before examining the specific 
influences of magnetic fields, it is
helpful to look briefly at the general energetic and dynamical
behaviors of the bubbles in the modeled ICM environment. 
For instance, the evolution of instabilities, magnetic fields and
circum-bubble flows depends on the history of bubble motions.
Hence, this section provides a brief overview of how bubble inflation
evolves, and it describes evolutionary features that
are common to many of the simulations that were performed.  This
background will assist in understanding the subsequent discussion of
the evolution of specific models calculated.
Figures \ref{bubstat1} and \ref{bubstat5c}  provide representative examples of the evolution of the bubble
energy contents and volume for very different energy injection times.
Figure \ref{bubstat1} illustrates model BM$_h$ (horizontal magnetic field
with constant $\beta$),
$t_i = 10~{\rm Myr}$. Other behavior of this model, to be discussed
below, can be seen in Figure \ref{HB1Tbd}. 
As explained already in \S 2.2, the bubble inflation power, $\dot E_b$
increased through the inflation period as the
bubble accelerates its expansion into the ICM.
The total energy injected (per kpc in the third, $x$ direction) was about 
$6.4\times 10^{56}{\rm erg}$.
For convenience, length, time and energy in the figures are expressed in
natural units of the simulations, with time in Myr, length in kpc
and mass corresponding to ICM inside 1 kpc$^3$. The natural
energy unit then becomes, $E_0 = 5.5\times 10^{55}$erg.

At the end of the inflation period, the 2D bubble volume, $V_{bub} \sim 28{\rm kpc}^2$,
so a characteristic radius of the bubble is $R\sim \sqrt{V_{bub}/\pi} \sim 3{\rm kpc}$.
As the bubble subsequently expanded adiabatically, its volume
increased by a factor $\approx 2.7$ relative to the value
at the end of the inflation period, while the total energy
inside the bubble dropped by a factor $\approx (1/2.7)^{2/3} = 0.52$
as one would then expect. In this case, as in all the others we computed, the
total energy in the bubble was dominated by thermal energy. We note
that the spike in the kinetic energy seen at $t = t_i$ corresponds
to termination of the bubble inflation. At that moment conditions inside the
inflation region were no longer held fixed, so that the inflation
region was effectively ``released'' into the ICM. It was buoyant, of course, so
it briefly surged upward within the bubble.

It is worth mentioning the evolution of magnetic energy
inside the bubble, illustrated in Figure \ref{bubstat1}.
Even though the bubble thermal energy decreased as expected during
expansion, the bubble magnetic energy increased, albeit irregularly.
A purely toroidal field, $B_{\phi}$  would have decreased inversely
with the radius of the expanding bubble in order to satisfy
magnetic flux conservation. Alternatively, a fully disordered
field in 2D would decrease as $\sqrt{\rho}$ during bubble expansion. 
Either condition would have lead to a constant bubble
magnetic energy in 2D. The observed global increase in the bubble
magnetic energy was a manifestation of strong, organized internal bubble mass 
circulations, also associated with some of the irregularities in bubble kinetic
energy content.

Figure \ref{bubstat5c} provides similar energetics information for
model BS-C, which has the same geometry as the above case but a lower
value of $\beta$.  In this case 
the bubble inflation region was maintained
for the full 65 Myr of the simulation, by which time the bubble
approached the top of the grid. Figure \ref{HB5Cbd} illustrates
the general appearance of the bubble in this model. The bubble inflation
power, $\dot E_b$,
reached an approximate steady state 
$\sim 2.6 E_0/{\rm Myr} = 1.7\times 10^{42}~{\rm erg/sec/kpc}$ 
after 20 Myr, which corresponds
to the formation epoch of the de Laval nozzle near $z \approx 25~{\rm kpc}$. 
We confirmed that the flow was transonic at that location with
$v \approx c_s \sim 3-4 c_{sI}$. The
total energy flux through the nozzle closely matched the inflation
energy flux seen in Figure \ref{bubstat5c}. Fluctuations seen in the latter
quantity resulted from instabilities in the flow near the nozzle. 
Smith \etal (1983) pointed out that there should be a critical
energy flux for bubble inflation, which in our 2D Cartesian geometry
would be $\dot E_{c} \approx P_I c_I h$. Below this energy flux the de Laval
nozzle is expected to pinch off the flows, so that they are unsteady,
while above this energy flux steady flows can be maintained. For our
conditions, $\dot E_c \approx 1.5 E_0/{\rm Myr}$, or about 57\% of the
inflation kinetic energy flux that developed in this simulation.

The overall evolution of bubbles in the ICM under these two extremes
of energy injection times proceeded much as expected from intuition.
The status of continuing physical connections between real cluster radio
bubbles and their parent AGNs is not unambiguous. Almost
by definition these objects do not display radio loud, jet AGN
connections, so it is simplest to assume that they are independent.
It may be premature to draw that conclusion generally, however, since
faint radio bridges have been seen in a couple of cases 
(MKW3S and A133, Young \etal 2004). We leave that issue open for now.

\begin{figure}
\epsscale{1.0}
\plotone{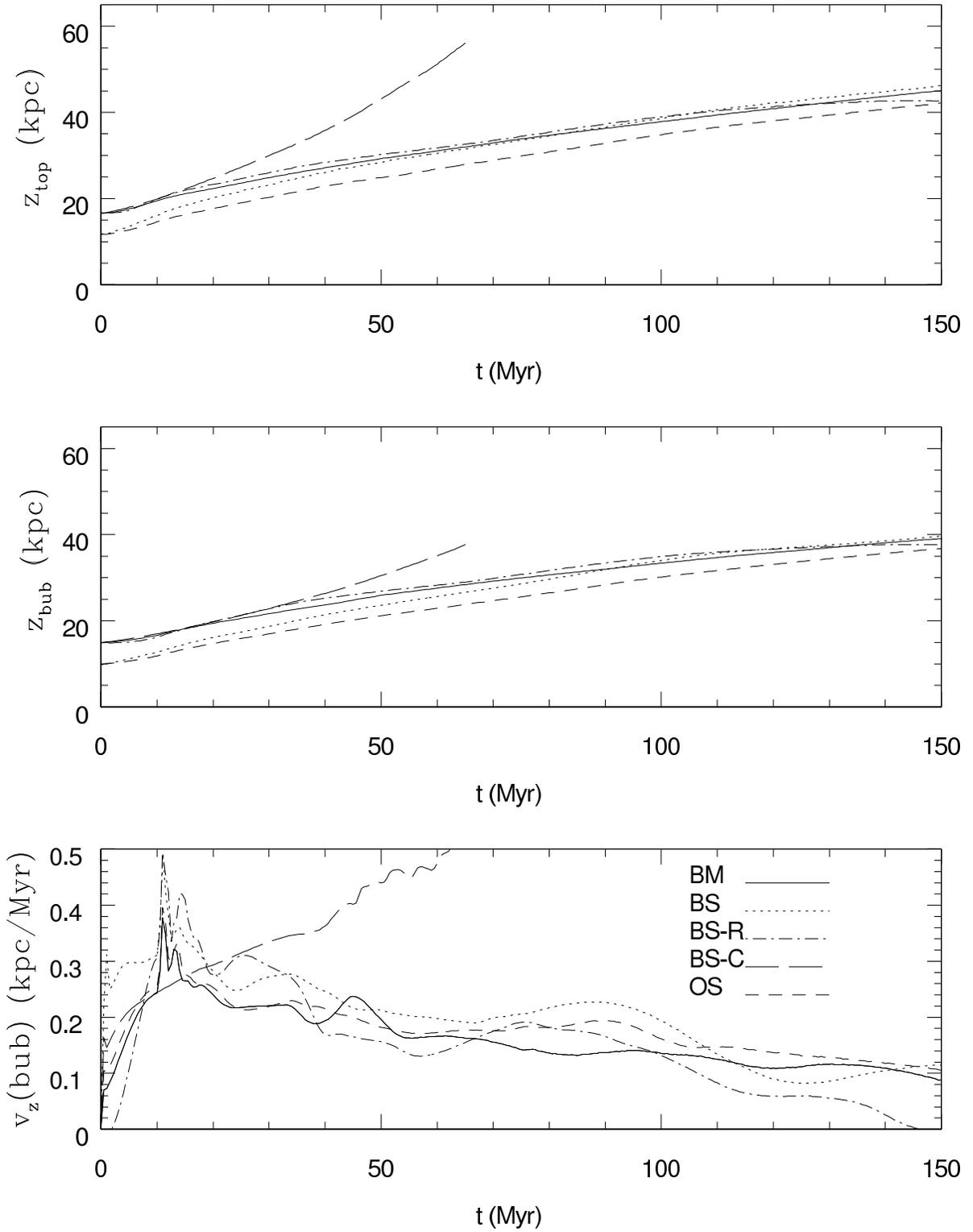}
\caption{Upward motions of several model bubbles as labelled. Top: position of
the top of the bubble; Middle: position of the mean height of the bubble;
Bottom: velocity of the mean bubble height.}
\label{bubdyn}
\end{figure}

\begin{figure}
\epsscale{.85}
\plotone{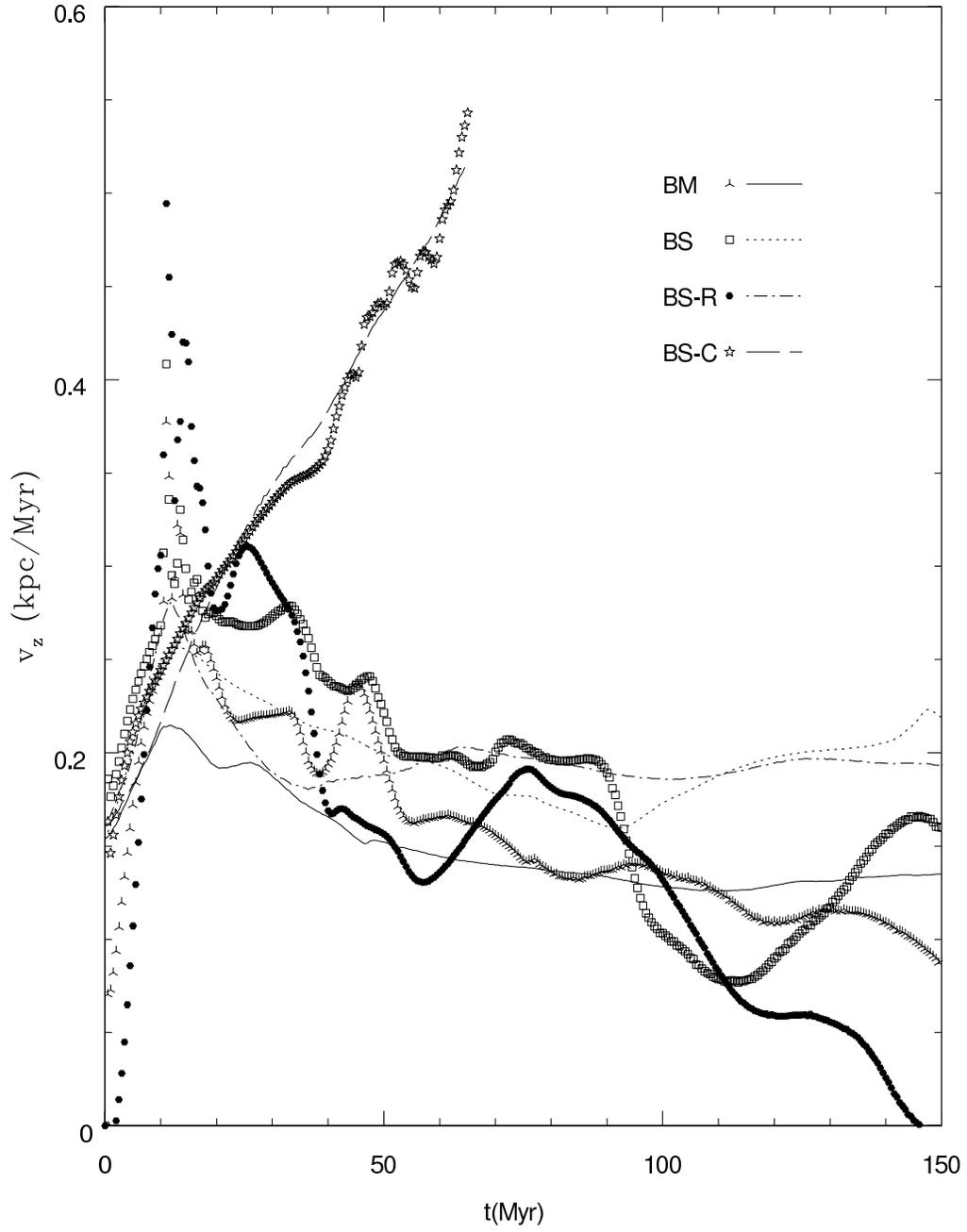}
\caption{Upward velocity of three model bubbles (points as labelled) compared to
a semianalytic buoyancy model as described in the text (equation \ref{uterm}).}
\label{uterm}
\end{figure}

Some additional key properties of bubble dynamics can also be gleaned from
Figures \ref{bubdyn}, \ref{uterm} and \ref{colvel}. Figure \ref{bubdyn}
illustrates the buoyant motions of several models. The
upper panel plots the position of the top of each bubble shown,
where $z_{top}$ is defined as the highest point at which the bubble
mass-fraction tracer $C_f \geq 0.95$. The center panel plots
the mean height of each bubble, defined as $z_{bub} = (\int z~C_f~dydz)/(\int C_f~dydz)$.
The bottom panel shows the mean upward velocity of each bubble,
defined as $v_z = d(z_{bub})/dt$. Except for model BS-C, which has ongoing
inflation,
all the models shown have the same short inflation period; namely, $t_i = 10~{\rm Myr}$.
The $v_z$ spikes at $t=t_i$ are caused by the upward surging
lower boundary of the bubble as it is released.
As we shall see, the various bubble models have rather different 
stability behaviors, morphologies and internal dynamics. Nonetheless, except for
BS-C, their buoyant upward motions through the ICM are qualitatively similar.

On the other hand, stability and circum-bubble flows depend essentially
on the
upward bubble velocity and also on the bubble acceleration.
To facilitate the relevant discussion it is helpful to derive
the usual expression for bubble terminal velocity, which is,
of course, determined by a balance between
buoyancy and drag. The general expression of the buoyant force is
\begin{equation}
F_{buoy} = -\int g~C_f~(\rho_{I} - \rho) dy dz.
\label{buoy1.eq}
\end{equation}
Substituting $-g = c^2_{sI}/h$, assuming a constant ICM
sound speed and that the bubble density
is small compared to the ICM, this can be approximated in a
convenient form
\begin{equation}
F_{buoy} \approx c^2_{sI} \left<\rho_{I}/h\right> V_{bub},
\label{buoy2.eq}
\end{equation}
where the angular brackets represent an average over the bubble volume,
$V_{bub}$.
The drag force can be expressed as
\begin{equation}
F_d = C_d L_y \rho_{I}(z_{top}) v^2_z,
\end{equation}
where $C_d$ is a drag coefficient, $\rho_{I}(z_{top})$ 
indicates the ICM density at $z_{top}$ and $L_y$ measures the
horizontal width of the bubble. Then the expected terminal velocity is
\begin{equation}
v_t = \sqrt{\frac{F_{buoy}}{C_d L_y \rho_{I}(z_{top})}}.
\label{uterm.eq}
\end{equation}
Figure \ref{uterm} compares $v_z$ (solid lines) 
for four model bubbles with $v_t$ found from equation \ref{uterm.eq}
using equation \ref{buoy2.eq} for $F_{buoy}$ (dashed curves). 
Here $L_y$ was set at each time to the
maximum y extent of the bubble gas, and it was assumed that 
$C_d = 2$ in each case. The simple, ``universal'' model, 
with constant $C_d$, represents the motions
of all the bubbles remarkably well. Obvious variances between the model
and the BS$_h$ and BS-R simulations late in the simulations resulted from MHD effects, as will be discussed
in \S 3.2.

The most important property of measured bubble motion after release was that it
generally decelerated. That resulted simply from the
fact that the scale height of the ICM increased with $z$, which, 
in turn, resulted from the decreasing gravity. We can see
this simply from equation \ref{uterm} if we approximate the bubble
cross section as a cylinder, so that
$L_y = \sqrt{4 V_{bub}/\pi}$. Since $V_{bub} \propto P_{I}^{-3/5}$,
and we pointed out in \S 2.1 the approximate ICM
behavior $P_I \propto z^{5/3}$, we expect from equation \ref{uterm}
that $v_z \propto z^{-1/4}$, roughly
consistent with the numerical results. This decelerated upward motion
reduced both the sheared flows around the bubbles and the strength of
vortical motions over time as they rose. That, in turn reduced disruptive
instabilities and the ability of the bubbles to lift ICM in
their wakes.

Representative examples of the velocity fields found in association with
the bubbles can be seen in Figure \ref{colvel}. The figure illustrates
the flows at the end of four simulations with short injection periods,
$t_i = 10~{\rm Myr}$. In each case ambient, ICM gas has been lifted in response to
vortical structures developed early on along the sides of the bubbles. 
As the vortical flows weakened and separated, returning
``down drafts'' developed underneath the bubble centers, as can be
seen in the figure. ICM was also pushed upward above
the bubble tops, especially in models with strong magnetic fields. Circulations
developed within the bubbles, as well. Their strength depends on the 
strength and geometry of the magnetic fields within the bubbles as
we shall discuss in the following subsection.

We comment that it is not reliable to use simulations with such limited
spatial domains to
examine quantitatively the redistribution of energy and entropy 
within the overall ICM (\eg Br\"uggen \& Kaiser 2002), since energy and gas do 
cross the outer boundaries of the computational domain. This effect will 
influence the large scale flows and modify the bulk transport of 
quantities.

\subsection{The Influence of Magnetic Field Strength and Geometry}

We begin our discussion of the
influence of magnetic fields with some brief general reminders about 
the instabilities
that largely determine the survivability of the bubbles, starting
with the Rayleigh-Taylor (R-T) instability.
The top center of a bubble is subject to the R-T
instability, especially early on, when the local
gravity is relatively stronger.  This will lead to dense fingers
of the ICM penetrating into the bubble from above. A tangential magnetic field
can stabilize the R-T instability (Chandrasekhar 1961) if the
restoring tension generated by bending of the field lines
exceeds the buoyancy force driving the instability. 
In the limit that the ICM density
greatly exceeds the bubble density and the magnetic field
strength continues across the contact discontinuity
the linear stabilizing condition can be
expressed approximately in terms of the MHD $\beta$ parameter as
$\beta < \beta_{crit} \approx 8\pi h/\gamma \lambda$, where $\lambda$ is 
the wavelength of the perturbation along the boundary (Jun \& Norman 1995).
Thus, the R-T instability is inhibited for small-scale surface perturbations
of wavelength $\lambda_{kpc} \la 25 h_{kpc}/\beta$, where
$h_{10kpc}$ expresses the scale height in kiloparsecs. 
In our simulations $h_{kpc} \approx 10$ where the bubbles form, while
initial $\beta$s range from $\beta \sim 10^2 ~-~10^5$.
In the weak field cases with $\beta > 10^3$, perturbations should initially be R-T 
unstable down to scales well below the size of the bubble.
On the other hand, we may expect the strong field cases, with
$\beta \sim 10^2$, to be R-T stable. These theoretical expectations 
are confirmed in the simulations as discussed below.

Disruptive Kelvin-Helmholtz (K-H) instabilities may develop in 
response to strong shear along the outer boundary of the bubbles once they
start to rise, when the ICM flows around them and 
strong vortices form along the bubbles in their wakes. In addition,
circulation developed inside the bubbles that contributed
strong shear at the bubble boundary. 
The strength and location of shear on the bubble boundaries
varied with time and with the magnetic field properties. Those details aside for
the moment, we find from inspection that tangential velocity differences
across the bubble boundary were commonly as large as 
$\sim 0.2~-~0.4~{\rm kpc/Myr}$ (that is, $\la 0.5 c_{sI}$) over much of the 
simulation time intervals,
even when the upward motion of the bubbles became less
than this late in the simulations (see Figures \ref{bubdyn}, \ref{uterm}).
The relatively large shear rates resulted in part because
the internal flow speed coming from circulation within
the bubble can be larger than the instantaneous upward speed of the bubble.

Tangential magnetic fields
can inhibit the K-H instability if the restoring magnetic tension induced by
bending the field lines exceeds the local ``Bernoulli lift force'' due to flows over
boundary corrugations. Formally, the instability is inhibited if the ``rms''
Alfv\'en speed across the boundary, $v_{A_{bnd}}$ (defined as $v_{A_{bnd}}^2
= (B^2_1 + B^2_{2})/(4\pi)(1/\rho_1 + 1/\rho_2)$),
exceeds the velocity difference across the boundary (Chandrasekhar 1961; Vikhlinin \etal 2001).
Since at the start of our simulations, the magnetic field just inside the bubble
boundary was typically at least as large as that immediately outside,
but the gas density was lower by two orders of magnitude, 
$v_{A_{bnd}}$ was
initially close to the Alfv\'en speed just inside the bubble
boundary. 

For discussion it is helpful to express $v_A$ in terms of the
sound speed and $\beta$; namely,
$v_A = \sqrt{\slantfrac{6}{5}} c_s/\sqrt{\beta}$. Using this expression
it is apparent that $v_A$ in the ICM 
was from one to two and a half orders of magnitude smaller
than the ICM sound speed for the models listed in Table \ref{tbl-1}.
At the same time, the bubble sound speed was larger by an order of
magnitude than $c_{sI}$, while $\beta$ inside the
bubble perimeter was initially smaller by about an order of magnitude.
Together, these produced an internal $v_A \sim v_{A_{bnd}}$ that 
initially ranged from being an order of
magnitude smaller than $c_{sI}$ in the weak field models to being
comparable to $c_{sI}$. 
Since the tangential velocity jumped across the
bubble boundaries approached a substantial fraction of $c_{sI}$ as
the bubble started to rise,
we expect the K-H instability to be operative initially in
the weak-field cases, but not in the strongest field cases. That is
confirmed by the simulation behaviors. 
The role of the internal magnetic field in more evolved bubbles
varies, depending on the strength and configuration of the initial
field, as discussed below.

\subsubsection{Short Bubble Inflation Periods}

\begin{figure}
{\bf See 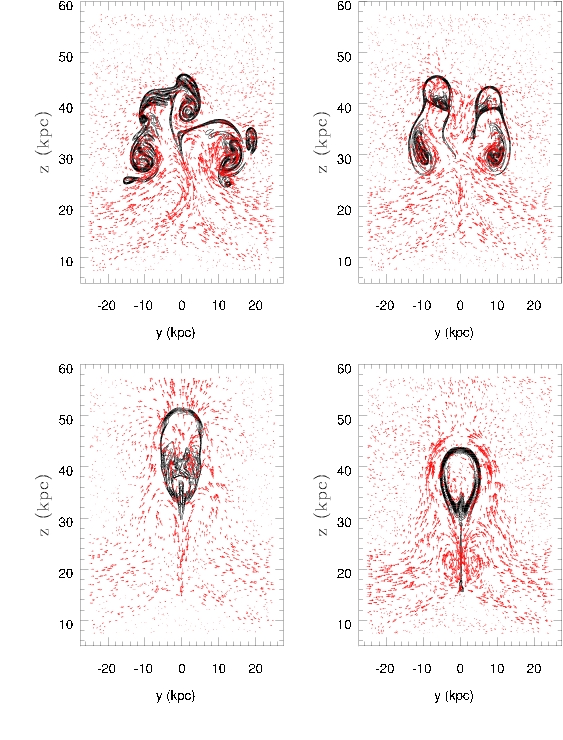}\\

{\bf High resolution image available at http://www.msi.umn.edu/Projects/twj/bubbles/f6.eps }\\

\caption{Velocity vectors overlaid on bubble mass-fraction ($C_f$) 
contours at $t = 150$Myr for
the BW (upper left), BM$_h$ (Upper right), BS$_h$ (lower left) 
and BS-R (lower right) models.  See the electronic version
of the Journal for a color version of this figure.}
\label{colvel}
\end{figure}

\begin{figure}
\epsscale{.85}
{\bf See 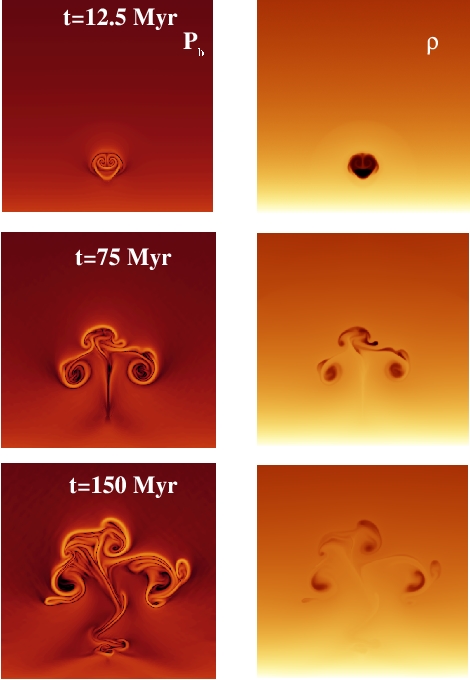}\\

{\bf High resolution image available at http://www.msi.umn.edu/Projects/twj/bubbles/f7.eps }\\

\caption{Snapshots from the evolution of the magnetic pressure ($P_b$)
and gas density ($\rho$) in the model BW 
($\beta_{I} \approx 7.6\times 10^4$). Images apply a
logarithmic color bar with high values represented by high tones.
The magnetic pressure spans six decades, with a peak value
$P_{b_{peak}} \approx 10^{-12}~{\rm dyne/cm}^2$ 
($B_{peak} \approx 5 \mu$G), while the gas density
spans 2.6 decades, with $\rho_{peak} \approx 7.6\times 10^{-25}~{\rm g/cm}^{-3}$. 
See the electronic edition 
of the Journal for a color version of this figure.}
\label{HBWSbd}
\end{figure}

\begin{figure}
\epsscale{.85}
{\bf See 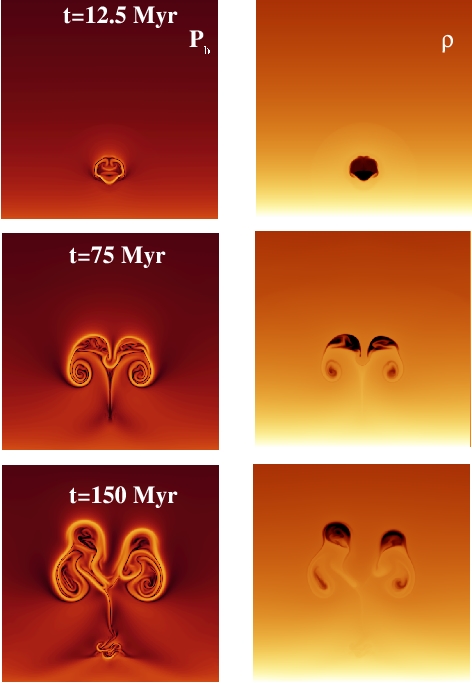}\\

{\bf High resolution image available at http://www.msi.umn.edu/Projects/twj/bubbles/f8.eps }\\

\caption{Same as Figure \ref{HBWSbd}, but for model BM$_h$
($\beta_{I} \approx 3\times 10^3$). The peak magnetic field reaches
about $20~\mu$G.}
\label{HB1Tbd}
\end{figure}

\begin{figure}
\epsscale{.85}
{\bf See 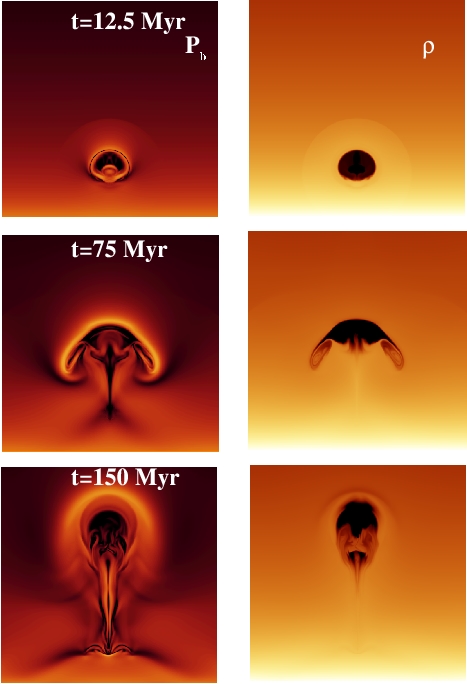}\\

{\bf High resolution image available at http://www.msi.umn.edu/Projects/twj/bubbles/f9.eps }\\

\caption{Same as Figure \ref{HB1Tbd}, but for model BS$_h$
($\beta_{I} \approx 120$). The peak magnetic field reaches about $40~\mu$G.}
\label{HB5kbd}
\end{figure}

\begin{figure}
\epsscale{.85}
{\bf See 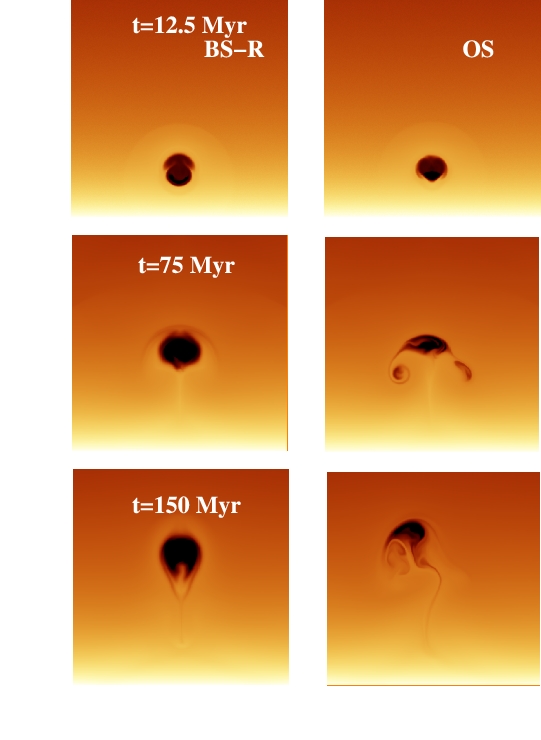}\\

{\bf High resolution image available at http://www.msi.umn.edu/Projects/twj/bubbles/f10.eps }\\

\caption{Snapshots from the density evolution of models BS-R and OS,
showing some of the dynamical influences of magnetic field geometry.
Display characteristics are the same as Figure \ref{HBWSbd}.
See the electronic edition
of the Journal for a color version of this figure.}
\label{fieldcomp}
\end{figure}

\begin{figure}
\epsscale{.85}
{\bf See 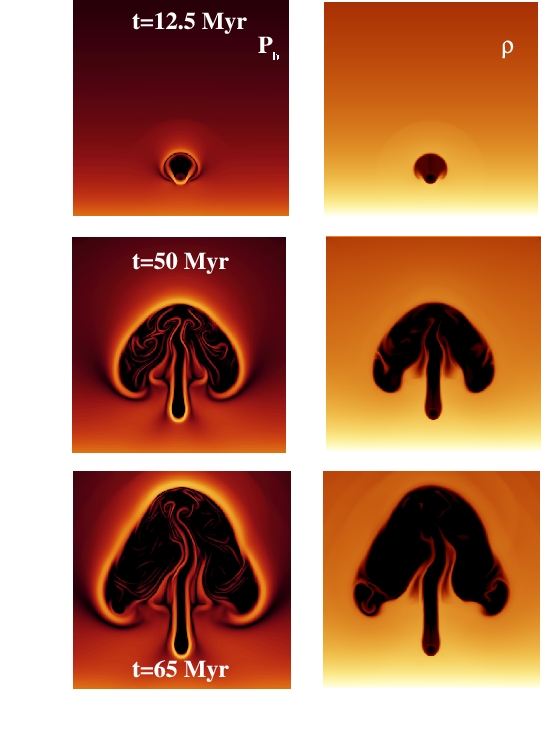}\\

{\bf High resolution image available at http://www.msi.umn.edu/Projects/twj/bubbles/f11.eps }\\

\caption{Magnetic pressure and density evolution for model BS-C,
which is the same as BS, except that $t_i = t_e$. See the electronic version
of the Journal for a color version of this figure.}
\label{HB5Cbd}
\end{figure}

General features of the evolutionary histories of three bubbles with 
an inflation period, $t_i = 10~{\rm Myr}$ are illustrated\footnote{mpeg animations
of these quantities can be found at\\
http://www.msi.umn.edu/Projects/twj/newsite/projects/bubbles/} in Figures \ref{HBWSbd}, \ref{HB1Tbd} and \ref{HB5kbd}. 
The images show
magnetic pressure and gas density structures for
bubbles inflated in constant $\beta_{I}$ atmospheres
with $\beta_{I} = 75550, ~3000, ~{\rm and}~120$.
The associated ambient field values at $z = 30~{\rm kpc}$ are 
$0.2,~1, ~{\rm and}~5~\mu$G, respectively. 
Near the base of the atmosphere the ambient field was roughly four
times the value at $z = 30~{\rm kpc}$, while it dropped to about
half that value near the top of the computed region.
The bubble plasma (as established by $C_f \approx 1$) was closely coincident with
the darkest tones in the density images. This provides
a relatively easy way to follow evolution of the bubbles.
In particular, their integrity 
can be ascertained by examining continuity of dark structures.
At $t = 150~{\rm Myr}$ the density distributions 
in Figures \ref{HBWSbd}, \ref{HB1Tbd}, \ref{HB5kbd} and \ref{fieldcomp}
can be directly compared to the mass-fraction contours in Figure \ref{colvel}.

The weakest field case (model BW, Figure \ref{HBWSbd}) with
$\beta_{I} = 75550$, corresponding to ICM fields of a few tenths of
a $\mu$G, evolves 
almost, but not quite, hydrodynamically. Thus, as expected, the bubble
was completely disrupted into fragments by R-T and K-H instabilities
soon after its inflation ended, and before it could rise significantly above 
the inflation region. One can see evidence of the R-T instability
along the bubble top and the K-T instability on
the sides in the $t = 12.5$Myr density image. By
$t = 75$Myr the bubble was largely twisted and stretched into
several barely connected fragments.
Eventually, even in this case, magnetic fields
played a stabilizing role in the behavior of bubble fragments, 
but too late to save the original bubble. The magnetic pressures
illustrated in Figure \ref{HBWSbd} reveal that ICM magnetic fields 
became wrapped around and stretched between the bubble fragments.
Generally, the field lines aligned with the most intense field structures,
so the highlighted features in the images approximately illustrate
the field topology in intense field regions, as well.
In this case fields were amplified locally to as much as several $\mu$G,
corresponding to Alfv\'en speeds, $v_A \sim 0.2~{\rm kpc/Myr}$, which
are comparable to the maximum velocity jumps found. 
Magnetic reconnection induced by circulation within the larger bubble fragments 
reduced the internal field after some initial amplification due to
stretching. Consequently, the enhanced ICM field was dynamically more important
than the internal field during later stages of the simulation.

Because the early evolution was essentially hydrodynamical, the 
behavior of this bubble is qualitatively
similar to the hydrodynamical bubble shown in Figure 1 of
R04. The most apparent difference between the density distributions of
the two simulations is the left-right
mirror symmetry in the R04 bubble, contrasting with the
asymmetric evolution of our BW bubble (resulting from small
breaks in the symmetry of the initial conditions) that make
its morphology more irregular.

A curious magnetic feature formed late in
the evolution of this bubble model as well as in all others
we have carried out that terminate bubble inflation well
before the end of the simulation. This feature is most obvious in the $t = 150$Myr
magnetic pressure image of Figure \ref{HBWSbd}, where magnetic sheets
can be seen extending down below the bubble material and folding over one 
another near the base of the ICM. Analogous
features can be seen in Figures \ref{HB1Tbd} and \ref{HB5kbd}.
This magnetic structure formed as magnetic fields were stretched inside
the downdraft
that forms as the upward motion of the bubbles decelerated. Counter-rotating
vortices that developed initially along the perimeters of the
rising bubbles dragged ICM material upward in the bubble wakes
and formed a current sheet near the center-line of the wake. This 
can be seen in the velocity fields of Figure \ref{colvel}. As the
upward motion of the bubbles decelerated, these vortical
patterns weakened and separated, allowing ICM material directly
underneath the bubble to settle again, depositing
the paired sheets of oppositely directed ambient magnetic field
separated by a thin current sheet. The folded sheets were
in the process of annihilating through reconnection, but that 
rate was roughly matched by the deposition rates in the simulations.

The model BM$_h$ ($\beta_{I} = 3000$) 
illustrated in Figure \ref{HB1Tbd} also began its evolution 
almost hydrodynamically.
Once again, R-T and K-H instabilities were clearly in evidence
during the early evolution, while strong
vortices developed along the bubble sides that incorporated significant
bubble plasma. The bubble plasma inside these two lateral vortices was 
eventually shed from the main bubble mass.
However, both external and internal magnetic fields were quickly strengthened as the
bubble inflated and began to rise. External field was 
pushed up with the rising bubble, then stretched and amplified around
the bubble and wrapped into the vortices. 
Inside the rising bubble
an additional pair of counter-rotating vortices formed during inflation that 
stretch the internal magnetic fields into intensified flux sheets along
the bubble walls. This further inhibited R-T and K-H instabilities on
the bubble perimeter. Near the top center, however, 
it is also evident in the $t = 75 {\rm Myr}$
images that the bubble was bifurcated by a dominant R-T finger formed
near the center of the bubble top. So, in this location, the R-T
instability continued. This behavior resulted from 
the initial magnetic reversal between the ICM and bubble fields on top of
the bubble combined with the effects of internal bubble motions. 
The initially anti-parallel fields on the bubble top 
reconnected in this vicinity. In addition, the internal counter-rotating
vortices deformed the internal magnetic field near the bubble top
into anti-parallel sheets that also reconnected (see, for example
the magnetic pressure images at $t = 12.5 {\rm Myr}$ in Figures \ref{HBWSbd}
and \ref{HB1Tbd}).
Consequently, the magnetic field was locally much reduced
compared to the rest of bubble perimeter, and this allowed the R-T
instability to continue in this area into the nonlinear phase.

To explore further the above physics behind the bifurcation of the B1-H bubble, 
we carried out a complimentary pair of simulations at half
the resolution of BM$_h$. The first of these, BM,
was identical to BM$_h$ except for the lower resolution and behaved in a 
fashion very consistent with the higher resolution case just described. 
The other simulation, BM-R, was identical to BM
except that the initial bubble magnetic field was 
reversed, so that it was parallel to the external field on the 
bubble top, but anti-parallel on the bubble bottom. 
As expected from the previous discussion, the
bubble in BM-R was not disrupted. Except for the
shed lateral vortices, which became magnetically isolated
from the main bubble body, it remained intact, since
the initial instabilities were quenched by magnetic tensions
over the entire top portion of the bubble. 
The same pair of internal vortices developed in the BM-R bubble,
and their motions still reduced the field inside
the bubble top. However, the amplified external field was
strong enough to inhibit nonlinear growth of the R-T instability.
In fact, as the upward motion of the bubble decelerated the main
bubble body developed a quasi-circular cross section in response
to the external magnetic hoop stresses.

After release from the inflation region, internal bubble circulations
in the BM bubbles amplified the field just inside the bubble wall 
until the local Alfv\'enic Mach number was of order unity, thus
damping the circulation there at later times.
Deeper inside the bubble fragments, however, circulation 
annihilated much of the magnetic field, so the motions continued.
By about $t \sim 40{\rm Myr}$ K-H instabilities
were largely inhibited by the strengthening surface field, 
although the R-T finger formed early on
continued to develop. 
The magnetic field on both sides of the
top bubble surface intensified to values as large as $20~\mu$G
as the bubbles rose, so that Alfv\'en speeds approached $c_{sI}$
and $\beta \rightarrow 10$.

The bubble in the stronger ambient field simulation, BS$_h$, 
($\beta_{I} = 120$; Figure \ref{HB5kbd}) was mostly stable 
against R-T and K-H instabilities from the very start, as predicted
in our simple analysis at the start of \S 3.2. 
The bubble remained intact.
On the other hand, just as for the BM and BM$_h$ models,
the magnetic field of the bubble and the ICM were anti-parallel on the 
bubble top and sides. There was significant reconnection in these
regions as the bubble started to push upward.
Similar to the BM cases, this produced a noticeably 
weakened magnetic field along the top and sides of the bubble.
On the other hand, the internal magnetic field quickly became
strong enough to inhibit formation of the internal, vortex pair
seen in the BM bubbles. This resulted in little internal magnetic
reconnection, and the R-T instability on the bubble top never developed.
However, a lateral vortex pair did form out of the K-H instability on
the trailing edges of this bubble, similar to the BM bubbles.  
In this case, however,
the magnetic field stretched inside those vortices quickly
became large enough to distort and disrupt the vortex flow. 
The local Alfv\'enic Mach number there dropped to near unity, as did
$\beta$. Previous MHD studies of the 2D K-H instability
have shown that such evolution leads to distended vortices similar
to those visible in Figure \ref{HB5kbd} (Jones \etal 1997).
Eventually, the lateral vortices were
pulled into the main bubble structure, squeezing it laterally and
actually accelerating its mean upward motion. Magnetic
fields on the top of bubble reached values almost an order of magnitude
larger than the initial field surrounding the bubble. The Alfv\'en
speed on both sides of the bubble boundary and deep into the
bubble interior approached $c_{sI}$, so that the magnetic fields
played a very important role in the motions of the bubble
and in the circulations within the bubble.
In particular, 
the magnetic hoop stress at the bubble top, $F_M \sim v^2_A \rho/R$,
where $R$ is the bubble radius, became comparable to the buoyancy 
force per
unit volume, $F_{buoy} \sim \rho_{I}|g| \sim c^2_s \rho_I/h$. Their
ratio is $F_M/F_{buoy} \sim (v^2_A/c^2_{sI})) (h/R)$.
We pointed out that $v_A \sim c_{sI}$. Since $R \sim h$,
this ratio was close to unity.
In fact the upward motion of the BS$_h$ bubble almost stalled before
the body of the bubble was briefly thrust upward after $t \sim 110 ~{\rm Myr}$ by the previously
mentioned interaction with the lateral vortices (see Figure \ref{uterm}).

The BS$_h$ model (Figure \ref{fieldcomp}) provides another example of
the possible importance of the magnetic field topology to the
bubble dynamics. This model is identical to BS$_h$ except
that it has half the numerical resolution and the bubble magnetic
field is reversed to make it parallel with the ICM magnetic field
on the bubble top. Thus, it represents a stronger-field comparison test
similar to the one already discussed for the BM simulations.
Consequently,
the only reconnection between the bubble and ICM fields took place
very near the bottom of the bubble at the
start of the simulation. Hence, the bubble
magnetic field remained largely isolated from the ambient ICM field. The
K-H instability was completely inhibited, so no vortical motions
developed that involve both bubble and ICM plasma. The bubble remained
intact, and $\beta$ within the bubble dropped to values as low as two
by the end of the simulation. The internal magnetic fields became
well-ordered, looping neatly parallel to the external bubble boundary,
since internal bubble circulation was largely damped out. 
The external magnetic field decelerated the upward motion of the 
bubble, similarly to model BS$_h$ (or the lower resolution equivalent, BS).
However, the BS-R bubble did not receive the late push from the lateral vortices,
so its upward motion stalled (see Figure \ref{bubdyn}). At the same
time, the bubble exhibited a breathing mode oscillation with a period
of about $50~{\rm Myr}$, which is
similar to the natural, Brunt-V\"ais\"al\"a period ($\sqrt{h/|g|}\sim 40 {\rm Myr}$).
Figure \ref{fieldcomp} illustrates the density evolution of model BS-R,
which can be compared to that of model BS$_h$ in Figure \ref{HB5kbd}.
The velocity fields at $t = 150 ~{\rm Myr}$ can be compared for the
two models in Figure \ref{colvel}. 

As one final illustration of the influences of magnetic field
geometry we also show in Figure \ref{fieldcomp} the density evolution
of model OS. In this model the ambient magnetic field was tilted, obliquely
at a 45 degree angle with respect to gravity.
Absent distortions around the bubble inflation region, the ambient field
would have been a uniform 5 $\mu$G, pointing down and to the right. Thus,
it was comparable in strength to the BM simulations in the
bubble inflation region, but to the BS simulations where the
bubble was found at $150~{\rm Myr}$. The evolution was something
of a hybrid between the two classes, with an added feature due to
the obliquity of the field. The bubble was initially subject to both
R-T and K-H instabilities. An R-T finger briefly began to develop
on top of the bubble (barely visible at $t = 12.5~{\rm Myr}$ in Figure \ref{fieldcomp},
but it was disrupted by the lateral stresses soon imposed on
the bubble by the external field. This inhibited the fragmentation
process that destroyed the bubbles formed in regions of
similar magnetic field strength for both horizontal and vertical
field geometries. Although the bubble became highly turbulent
it remained intact to the end of the simulation.

\subsubsection{Continuous Inflation}

All the simulations discussed above involved bubble inflation periods
that were short compared to the time required for the bubble to rise
through a scale height in the ICM. Our simulations suggest that so long 
as this condition is satisfied the subsequent evolution
will be qualitatively independent of the inflation period. This conclusion
is apparent from the qualitative similarities between our results 
discussed so far and from the simulations of R04, which used only preformed
bubbles. For our ICM and bubble characteristics the critical inflation
timescale is about $20~{\rm Myr}$. Br\"uggen \etal (2002) 
and Br\"uggen \& Kaiser (2002) presented results of hydrodynamical
simulations in which a bubble inflation mechanism similar to ours
was kept constant. Their unmagnetized, buoyant plasma developed flocculent
plumes, so were rather different from the briefly inflated MHD bubbles 
discussed above or the preformed MHD bubbles simulated by R04. 

In order to investigate the MHD properties of continuously inflated
bubbles we carried out two simulations, BS-C and BW-C, described in
Table \ref{tbl-1}.
We found qualitatively similar structures to the HD plumes of
Br\"uggen \& Kaiser (2002) for our constantly inflated
MHD bubbles, as illustrated
in Figure \ref{HB5Cbd}. As pointed out in \S 3.1, the continuously
inflated hot bubble formed a de Laval nozzle about one scale height above
the inflation region, with mildly supersonic flows ($M\sim 1.1$) 
inflating the bubble body
at higher altitudes. The plume was highly unstable to bending modes
(in 2D, planar geometry), so that flows
internal to the bubble, while subsonic, were chaotic with speeds several times $c_{sI}$.
The bubble shown, BS-C, was formed in the same ICM as bubble
BS$_h$, discussed above. The ICM was given an initial $\beta_{I} = 120$.
The bubble was largely stable to disruptive R-T and K-H 
instabilities, like others with the same ambient magnetic field.  
As before, the ambient magnetic field was pulled up with the bubble and
became roughly as strong along the bubble perimeter as it was
in the BS$_h$ simulation. Because of the strong and chaotic internal
circulations within the bubble, the bubble magnetic field
became highly filamented, but generally weak, due to magnetic reconnection.

We also carried out an analogous simulation, BW-C, with continuous inflation,
but with the same weak
magnetic field conditions as BW ($\beta_{I} = 75550$). This
bubble remained essentially hydrodynamical until the end of the
simulation, when it crossed the upper grid boundary. On the other
hand, its evolution was qualitatively very similar to the B5-C bubble, 
except that R-T and K-H instabilities caused the bubble boundary to
become very flocculent. Unlike the ``free-floating'' bubble BW, however,
the Alfv\'en speed nowhere became comparable to the flow speeds
along the bubble boundaries, or within the strong, internal bubble
circulations. Also, in this case the flows above the de Laval nozzle
reached Mach numbers approaching $M=2$, leading to an intermittent 
terminal shock where the rising plume entered the bubble, as well as weak
shocks within the bubble itself.

\section{Summary and Conclusions}

We have carried out an extensive set of 2D MHD simulations of hot
buoyant MHD bubbles in a realistic ICM. Our bubbles were inflated
over a finite time near the base of the ICM from a region of circular 
cross section and with uniform gas pressure and density and a circumferential 
magnetic field initially isolated from the ICM magnetic field. 
The ICM magnetic fields covered a range of
strengths and geometries. The simulations support
the basic findings of previous analytic studies and 
2D simulations and add substantial new
insights. As expected from previous work, absent or 
very weak fields permit development of R-T and K-H instabilities,
so that bubbles fragment quickly upon release into the ICM.
On the other hand, as expected, moderately strong ICM magnetic fields
of a few $\mu$G can stabilize R-T and K-H instabilities along
the surfaces of the bubbles as they start to rise. Such bubbles
remain intact as they rise through several scale heights in the ICM. 
Magnetic flux from the ICM is pulled up and over the rising 
bubble while being wrapped into
vortices that form along the sides and in the wake of the bubbles.
Those motions stretch and strengthen the ambient field, so that
its role becomes more important as the bubbles evolve. Even though
the initial magnetic pressures in our simulated environments were
all less than 1\% of the ambient gas pressure and the Alfv\'en speeds
at least an order of magnitude less than the sound speeds, 
the magnetic pressure surrounding
the bubbles can become a significant fraction of the gas pressure and 
the Alfv\'en speed comparable to the ICM sound speed and bulk flow
speeds. Magnetic tension
from the field draped over the bubbles can then help retard their
upward movement and influence the morphology of the bubbles.  
Following their amplification around a bubble, even 
very weak ambient magnetic fields (less than a $\mu$G) 
eventually add stability to the bubbles or their fragments. 

In addition our simulations show that the geometry of the field and
the field topology at the interface between the bubble and the ICM
play significant roles in the evolution of the bubbles and their
fragments. For instance, we find sometimes strongly different 
behaviors arising, depending on
whether the initial bubble and ICM fields were parallel or 
anti-parallel on the bubble top. In particular, anti-parallel fields 
there tend to reconnect, the reducing field strengths and their ability to
inhibit disruptive instabilities. The orientation of the ICM field with
respect to the direction of gravity also is significant. It is already
well-known in 2D models that magnetic fields orthogonal to the
computational plane have limited effect, since they contribute only a
pressure force. Previous simulations have demonstrated differences
between vertical and horizontal ICM fields. Vertical fields have 
limited influence on initial bubble stability, but can help confine its 
lateral expansion, while horizontal fields influence both behaviors.
We looked, in addition, at the evolution of bubbles in oblique
magnetic fields. Their roles are more comparable to those of
horizontal fields in that they inhibit initial instabilities at a
reduced level compared to a horizontal field of similar strength,
since field lines become draped around a rising bubble. 
On the other hand, the Maxwell stresses induced
in this situation are decidedly oblique as well, so they contribute to
shearing motions within the rising bubbles that can help disrupt the
nonlinear R-T fingers that do form.

The dynamical roles of the magnetic fields internal to the bubbles
are more complex. Strong internal fields parallel to the 
boundary
can play similar roles to the external field, of course.  Such fields
can develop in response to vortices
driven by differential and time varying buoyancy within our bubbles. 
In general, rising bubbles are subject to strong, varying and 
irregular stresses, so in many cases they develop internal 
circulations with speeds approaching
the upward velocity of the bubble, unless the magnetic fields
are strong enough to inhibit them; that is, unless the Alfv\'enic
Mach number of the flows is relatively small. The internal motions
can lead to locally strong fields where they are stretched, but
they also lead to substantial field annihilation inside vortices. Consequently,
the growth of total magnetic energy inside the bubbles is generally modest
and irregular.

ICM gas and magnetic fields are, of course, dredged up in the bubble wakes
in response to vortical flows that developed behind the bubbles. 
The history of those motions reflects the rising motions of the
bubbles themselves. The upward velocities of our simulated bubbles 
are well-described by simple terminal velocity models balancing buoyancy
against ram pressure drag with a constant drag coefficient appropriate
to a blunt object. During the inflation process the terminal bubble velocities
typically increase over time, since they rapidly
displace more and more dense, ICM material. On the other hand,
once inflation ends, our bubble terminal velocities generally
decrease, due to the weakened gravity at higher altitudes.
This effect, which seems very likely to apply in most real clusters,
has a significant effect on the subsequent evolution of the bubbles.
As a bubble decelerates, its wake vortices weaken and
separate. In response a return flow develops in the ICM below the bubble
that carries magnetic flux with it. In our 2D
geometry a thin current sheet forms between oppositely directed
fields. In addition, as the bubble decelerates the gravitational
and Reynolds stresses that contribute to instability development
are reduced. Hence, remaining bubble fragments are able to relax, and
any surrounding amplified fields formed during earlier motions
are also more effective in bubble stabilization.

We also explored the influence of varying the length of time over
which the bubble is inflated. While short inflation periods produce
behaviors similar to those of preformed bubbles, inflation over long 
periods leads to mushroom-capped rising plumes with internal
supersonic motions and, in some cases, intermittent internal shocks.
The transition between these behaviors
depends on whether the bubble inflation is less than or greater than
the time for buoyant material to rise through an ICM scale height
above the inflation region. Since buoyant velocities are a significant
fraction of the ICM sound speed, the critical inflation period is
of the order of the ICM sound crossing time. 

\subsection{Astrophysical Implications}

Several conclusions that are relevant to the formation and evolution
of relic radio bubbles in clusters of galaxies can be drawn from these
MHD simulations.  Some of these are quite straightforward, while
others are more complex or will require additional effort for their
complete understanding. However, the principal points of astrophysical
interest are as follows:

1) First and foremost, it is now very clear that intracluster magnetic
fields that are initially dynamically unimportant can stabilize the
relic radio bubbles against disruption by surface instabilities in the
form of Rayleigh-Taylor and Kelvin-Helmholtz processes. Though there
is currently some uncertainty in the exact values for many ICM magnetic
fields (e.g., Carilli \& Taylor 2002, Taylor et al. 2002, Rudnick \& Blundell 2003), there
is broad agreement that fields of $1-10 \mu$G are most likely. 
The
resulting values of $\beta$ for the ICM then lie in the range of a few
tens to a few thousand, depending on the richness of the cluster and the
location within the cluster. For such values of $\beta$ the present
calculations then show, in agreement with previous estimates, that the
radio bubbles are stablized against disruption.

2) A second and related conclusion involves the buoyant risetimes of
the bubbles. These times have been estimated by Birzan et al. (2004)
from the observational data to lie in the range from a few tens of 
millions of years to over $10^{8}$ years, and a similar value of
$\sim 10^{8}$ yr was obtained by Churazov et al. (2001). 
The calculations presented
here, which include the inflation of the bubbles and realistic
ICM and gravitaty conditions, are consistent with
these estimates.  A particular feature of interest in the dynamical
history of the bubbles is the deceleration they experience and the
role played by the magnetic field in this process, as described in
Section 3.  It would be of great interest in determining the ultimate
fate of relic radio bubbles to follow their late time evolution in
the ICM; this calculation will have to await a treatment in three
dimensions.

3) A key issue in many observational and theoretical discussions of
relic radio bubbles is their potential role in reheating cluster cooling
flows, either by direct injection of energy or by mixing regions of
the ICM that are at different temperatures. These simulations provide
strong evidence of mixing and lifting of the ICM, together with mixing
of the bubble material with the ICM.  The location of these mixing
regions can be along the sides of the bubble, in the bubble wake, and
even along the top surface of the bubble.  However, as discussed in
Section 3, the amount of lifting and mixing is a strong function of
the ICM, $\beta$ and the relative orientation of the ICM and bubble
magnetic fields.  Magnetic field reconnection plays an important role
in mediating this mixing process, and it emphasizes the need for an
MHD treatment. In general, mixing and lifting is most pronounced in
the wake of those bubbles that more closely resemble the observational
data; i.e., for those bubbles with stable geometries and relatively
low values of $\beta_{I}$. At present it is not possible to provide
a quantitative estimate of the overall mixing and lifting in the ICM
from bubble evolution due to the limitations of the overall simulation
volume, as discussed at the end of Section 3.1. Thus, while relic radio
bubbles clearly contribute to reheating of the ICM, a quantitative
assessment of this effect requires additional, carefully designed calculations.

4) The discussion of bubble dynamics in Section 3.2 also illustrates
the importance of the interior dynamics and magnetic field structure
of the bubbles in determining their interaction with the ICM. The
bubbles simulated here have very simple properties; they are filled
with hot, rarefied gas and have well defined field geometries. Yet, the
subsequent evolution of their interior structure is complex, and it
is hoped that future three-dimensional simulations will provide enough
detail that inferences can be drawn about the internal composition of
the bubble when comparisons are made with observations. This is a
question of particular interest for two reasons.  First, the observational
data seem to show that the internal energy of the bubbles exceeds their
minimum equipartition energy obtained from radio data by factors of 
10 or more (McNamara et al. 2002, Birzan et al. 2004), and second,
the particle content of radio jets is still unknown and remains one of
the critical outstanding problems in this field. Hence, any insights
into the particle content of the radio bubbles would be of great interest.

\subsection{Future Work} 

A full understanding of MHD influences on the dynamics of
buoyant bubbles in cluster media must await high resolution 3D MHD
studies. We can, however, anticipate some of the important
similarities and differences. Two important limitations of 2D
flows may be important to keep in mind. First, there is no vortex stretching
in 2D flows, which in 3D leads to significant strengthening of flux
ropes. Thus, in 3D we may expect more intense local field amplification
in complex flow regions. Our 3D MHD simulations of light, 
supersonic jets penetrating ICMs are consistent with that
expectation (\eg Tregillis \etal 2001; O'Neill 2004), as are comparisons
between our 2D and 3D simulations of the MHD K-H instability (\eg
Jones \etal (1997); Ryu \etal (2000)). On the other hand, the
relative global magnetic field enhancements seen in the 3D jet
simulations are once again modest and qualitatively consistent
with our results here in an admittedly somewhat different flow.
Both, however, are driven flows that have generally chaotic
behaviors.

One of the dominant MHD behaviors seen in the present 2D MHD simulations
is the formation of dynamically strong field regions on the leading
surface of the rising bubbles. Analogous features have been 
reported in 2D MHD simulations of fast, dense clouds moving
through magnetized media (\eg Miniati \etal (1999)). One possible
difference in a 3D flow is the possibility that the magnetic flux
will 'slide' off the top of the rising bubble. Unless the bubble
is magnetically supported, that behavior is unlikely, however, as 
illustrated in the 3D MHD moving cloud simulations of Gregori \etal (1999).
They found that magnetic flux of the ambient medium tends to become
trapped in surface irregularities of the cloud, so that it is
indeed stretched around the perimeter. Depending on the local
field geometry and coherence length the dynamical outcome of that
behavior is not clear, however. While it stabilizes K-H instabilities
with wave vectors parallel to the field it does not do so in 
orthogonal directions. That behavior requires full 3D simulation to
be understood.

\acknowledgments

This work was supported at the University of Minnesota by NSF grant AST03-07600
and by the University of Minnesota Supercomputing Institute. The National
Optical Astronomy Observatory is operated by AURA Inc. under a cooperative
agreement with the National Science Foundation. Portions of this work were
performed during the 2004 astrophysics program at the Aspen Center for
Physics. We thank the referee for constructive comments on improving
the manuscript.

\clearpage

\begin{deluxetable}{ccrrrrrrrrcrl}
\tabletypesize{\scriptsize}
\tablecaption{Summary of Models Discussed\label{tbl-1}}
\tablewidth{0pt}
\tablehead{
\colhead{Model\tablenotemark{a}} & \colhead{Resolution\tablenotemark{b}} & \colhead{Ambient Field} 
& \colhead{$\beta_{I}$\tablenotemark{c}} 
& \colhead{$B_{I}$\tablenotemark{c}~\tablenotemark{e}}
& \colhead{$B_{bub}$\tablenotemark{f}~\tablenotemark{g}} 
& \colhead{$t_i$} & \colhead{$t_e$ } &\\
& & \colhead{Geometry\tablenotemark{c}~\tablenotemark{d}} & & \colhead{$\mu$G} & \colhead{$\mu$G} & \colhead{Myr} & \colhead{Myr} 
}
\startdata
BW & $512\times 512$ & hb & 75550 & 0.2 & 2(CCW) & 10 & 150\\
BW-C & $512\times 512$ & hb & 75550 & 0.2 & 2(CCW) & 75 & 75\\
BM &$512\times 512$ &hb &3000 &1 &10(CCW) &10 &150\\
BM$_h$ &$1024\times 1024$ &hb &3000 &1 &10(CCW) &10 &150\\
BM-R &$512\times 512$ &hb &120 &1 &10(CW) & 10 & 150\\
BS &$512\times 512$ &hb &120 &5 &50(CCW) & 10 & 150\\
BS$_h$ &$1024\times 1024$ &hb &120 &5 &50(CCW) & 10 & 150\\
BS-R &$512\times 512$ &hb &120 &5 &50(CW) & 10 & 150\\
BS-C &$512\times 512$ &hb &120 &5 &50(CCW) & 65 & 65\\
OS &$512\times 512$ & ou & 50 - 3000 & 5 & 10(CCW) & 10 & 110\\
\enddata


\tablenotetext{a}{See note below for key to model labels.}
\tablenotetext{b}{All simulations used a Cartesian domain $y = [-27.5,27.5]~{\rm kpc}$,
$z = [5,60]~{\rm kpc}$.}
\tablenotetext{c}{Measured away from bubble influence}
\tablenotetext{d}{ h = horizontal ($\phi_0 = 0\degr$), 
o = oblique ($\phi_0 = -45\degr$), v = vertical ($\phi_0 = 90\degr$);
b = constant $\beta_{I}$, u = ``uniform''} 
\tablenotetext{e}{z = 30 kpc}
\tablenotetext{f}{Maximum value measured at $r=r^{'}_b$ during bubble inflation}
\tablenotetext{g}{CCW = counterclockwise field orientation; CW = clockwise (or ``reversed'') field orientation}
\tablecomments{The primary two letter model designation indicates 
the geometry of the ICM magnetic field (column 3) followed by a qualitative
indicator of relative field strength; weak (W), medium (M) or strong (S).
The two high resolution simulations are additionally marked by the
subscript h. Continuous injection models are tagged with ``-C'',
while those with the bubble magnetic field reversed (CW) have the designation
``-R''.}

\end{deluxetable}

\end{document}